\documentclass[final,5p,times,twocolumn]{elsarticle}
\usepackage{macros}
\usepackage{changepage}
\newcounter{explanations}

\begin{document}
\begin{frontmatter}
\title{An Ontology for Representing Curriculum and Learning Material}
\author[wsu]{Antrea Christou}
\ead{christou.2@wright.edu}
\author[wsu]{Chris Davis Jaldi}
\ead{jaldi.2@wright.edu}
\author[ksu]{Joseph Zalewski}
\ead{jzalewski@ksu.edu}
\author[ksu]{Hande Küçük McGinty}
\ead{hande@ksu.edu}
\author[ksu]{Pascal Hitzler}
\ead{hitzler@ksu.edu}
\author[wsu]{Cogan Shimizu}
\ead{cogan.shimizu@wright.edu}
\address[wsu]{Wright State University, USA}
\address[ksu]{Kansas State University, USA}
\begin{abstract}
    Educational, learning, and training materials have become extremely commonplace across the Internet. Yet, they frequently remain disconnected from each other, fall into platform silos, and so on. One way to overcome this is to provide a mechanism to integrate the material and provide cross-links across topics.
    In this paper, we present the Curriculum KG Ontology, which we use as a framework for the dense interlinking of educational materials, by first starting with organizational and broad pedagogical principles. We provide a materialized graph for the Prototype Open Knowledge Network use-case, and validate it using competency questions sourced from domain experts and educators.
\end{abstract}
\begin{keyword}
Education \sep Ontology \sep Knowledge Graph \sep OWL
\end{keyword}
\end{frontmatter}
\section{Introduction}
\label{sec:intro}
The Internet and various platforms, from social media (e.g., Medium~\cite{site:medium}) to educational sites (e.g., Coursera~\cite{site:coursera}), have enabled a vast proliferation of educational materials across nearly any conceivable domain. However, this has resulted in various learning materials being scattered across different sites or platforms; becoming difficult to find or access; often redundant (or worse, contradictory), silo-ed away from other complementary material, or may be created once for a narrow use case and never used again. Moreover, educational materials span various formats, and thus multimodality remains a challenge.

Our goal is to provide a mechanism by which educational materials (across various forms and formats, e.g., multimedia) can be densely interlinked, so as to provide seamless progression across learning pathways, irrespective of the source materials' location. 

In the end, this is a classic scenario for a data integration task. Of course, Knowledge Graphs (KGs) are well-suited for this task \cite{kgs-hogan,FOST,kgs-noy}. Specifically, we envision a 
fully materialized RDF \cite{rdf-tr} graph equipped with Web Ontology Language (OWL) \cite{owl2-primer} ontology as a schema, which describes the metadata, content, and links between educational, learning, and training (ELT) material and, more broadly, curricula.

As such, this ontology is one first step towards reasonable personalized instruction, where the needs of the learner can be accounted for, alongside contextualization of each piece of ELT material. Indeed, by using formally structured organization (i.e., the ontology) for the metadata of the ELT material inventories, coupled with its interconnections, we make this an effective retrieval augmented generation (RAG) \cite{lewis-rag} target for agentic Large Language Models (LLMs) \cite{wang2024survey,singh2025agentic}. By emphasizing the agentic nature of these systems (e.g., using SPARQL \cite{prud2006sparql} to retrieve relevant information, which is in turn used to access or otherwise locate specific relevant documents), we furthermore facilitate an \emph{explainable} process, which is paramount in educational scenarios. Thus, in this paper, we lay the foundation for a class of agentic LLM applications, which we would call neurosymbolic pedagogical agents \cite{jws-napas}. Specifically and concretely, we cover in this paper our additions to the state of the art:
\begin{compactitem}
  \item the Curriculum KG Ontology, specified in OWL 2 DL \cite{owl2-primer};
  \item its materialization, serialized in RDF \cite{rdf-tr}; and
  \item its validation and a worked example, via competency questions.
\end{compactitem}

Next, Section~\ref{sec:description} describes the ontology for the Curriculum KG, with its materialization and evaluation expanded towards Section~\ref{sec:kg}. 
This is followed by a brief discussion of related work in Section~\ref{sec:relatedwork}, accompanied by a brief critical analysis on how our work advances the state of the art.
Finally, in Section~\ref{sec:conc}, we conclude with next steps.

\section{The Curriculum KG Ontology}
\label{sec:description}
In this section, we discuss the design methodology, overarching use-case, and formalization of the Curriculum KG Ontology. The formalization and accompanying documentation can be found online \cite{repo:currkg}. The ontology and KG (Section~\ref{sec:kg}) are licensed under a dual license for complete coverage: the ontology, schema diagrams, and the KG itself are under the CC-BY-SA-4.0 license, whereas the materialization scripts are licensed under the Apache 2.0.
We note that the materials themselves retain their own licenses (and, indeed, they are modeled -- and indicated -- as such in KG).

\subsection{Use-case scenario}
\label{ssec:scenario}
The Prototype Open Knowledge Network (Proto-OKN) \cite{site:proto-okn} is a collaborative program across six U.S. Federal agencies, including the National Science Foundation (NSF\footnote{\url{https://www.nsf.gov/}}), the National Aeronautics and Space Administration (NASA\footnote{\url{https://www.nasa.gov/}}), the National Institutes of Health (NIH\footnote{\url{https://www.nih.gov/}}), the National Institute of Justice (NIJ\footnote{\url{https://nij.ojp.gov/}}), the National Oceanic and Atmospheric Administration (NOAA\footnote{\url{https://www.noaa.gov/}}), and the  U.S. Geological Survey (USGS\footnote{\url{https://www.usgs.gov/}}). The purpose is to build \emph{open knowledge network}, defined as ``publicly accessible, interconnected sets of data repositories and associated KGs that enable data-driven, artificial intelligence-based solutions for a broad set of societal and economic challenges.'' The program is organized into 15 projects that solve various domain-specific problems, two projects that provide the common infrastructure that integrates them, and one project that seeks to provide the educational, learning, and training interface to accessing, using, and leveraging the Proto-OKN. 

The Curriculum Knowledge Graph (CurrKG) and its ontology directly inform this final project: The Education Gateway for the Proto-OKN (EduGate), and subsequently provides the overarching use-case scenario for the design decisions and competency questions. For example, the CurrKG could be used to answer such questions as:
\begin{compactitem}
    \item What is the OKN and how can it be used to address climate change?
    \item What technologies do I need to understand and contribute to the OKN Project X?
    \item What are all the materials in the CurrKG that explain SPARQL?
    \item What are all the topics that have the most associated media resources in CurrKG?
\end{compactitem}
While the first two questions are specific to the Proto-OKN, it is the latter two which are of particular interest for broader impacts, including educational initiatives. We provide more detailed questions -- and their respective SPARQL queries -- in Section~\ref{ssec:eval}.


In the end, the primary objective, through the design, implementation, and deployment of the CurrKG ontology, will be to provide a model that supports the meaningful discovery of ELT materials and their contexts. Specifically, EduGate -- and through the CurrKG and its ontology -- seeks to model audience-specific curriculum, which we call a \textsf{Persona}. As such, we see this directly modeled in our ontology, and it becomes a guiding principle for our ontology design decisions.

\subsection{Design Methodology}
\label{ssec:design}
Individualized instruction generally requires significant flexibility. As such, we have chosen the Modular Ontology Design methodology \cite{momo-swj}.

Modular Ontology Modeling (MOMo) builds complex ontologies out of small, manageable subunits, which we call \textsf{Modules}. A module typically has one core concept and several ancillary concepts and properties conceptually related to its core concept, but may not be directly related to other modules. This segmentation facilitates understanding, future modifications to ontology (as each modification can often be confined to one module, making re-use and adaptation simpler \cite{enslaved-jws}, and can help with future (knowledge) alignment activities \cite{llms-alignment-kgswc,llms4kgoe-jws}. During the execution of MOMo for the CurrKG ontology, we created three modules: \textsf{Persona}, \textsf{Learning Path}, and \textsf{Module}.\footnote{We accidentally overload this term, but here we mean the concept of \emph{learning} modules -- as largely self-contained sets of educational material, and they should not to be confused with ontology modules. These are also called \emph{learning units} in some literature \cite{cs2023}.}

Often modules are designed by \emph{instantiating}\footnote{This is fully known as \emph{template-based instantiation}, from \cite{template}.} an \emph{ontology design pattern} (ODP) \cite{HGJKP2016}, a small generic ontology that reflects a modeling best practice. ODPs are also used in ontology design methodologies besides MOMo \cite{BlomqvistHP16}, and there are several online repositories \cite{site:odpa} and libraries \cite{modl} of ODPs. Our \textsf{Persona} module is based on the AgentRole pattern, and our \textsf{Learning Path} on the Sequence pattern, from the MODL library \cite{modl}.

In accordance with the final steps of the MOMo process, we annotated the ontology using the Ontology Pattern Language (OPLa) \cite{OPLa}, which makes the division of classes and properties into modules explicit.

\subsection{Overview of Concepts and Relations}
\label{ssec:overview}
This section is a non-exhaustive description of the CurrKG ontology; for brevity, we have not specified lists of subclasses, all data properties, or the contents of controlled vocabularies. For complete documentation, see \cite{repo:currkg}. 
Our ontology is formalized in the OWL 2~\cite{owl2-primer,owl2-semantics}, specifically in OWL 2 DL. Below, selected axioms are shown in each respective block of text, expressed using DL notation.

\begin{figure*}[p]
    \centering
    \includegraphics[width=1.0\linewidth]{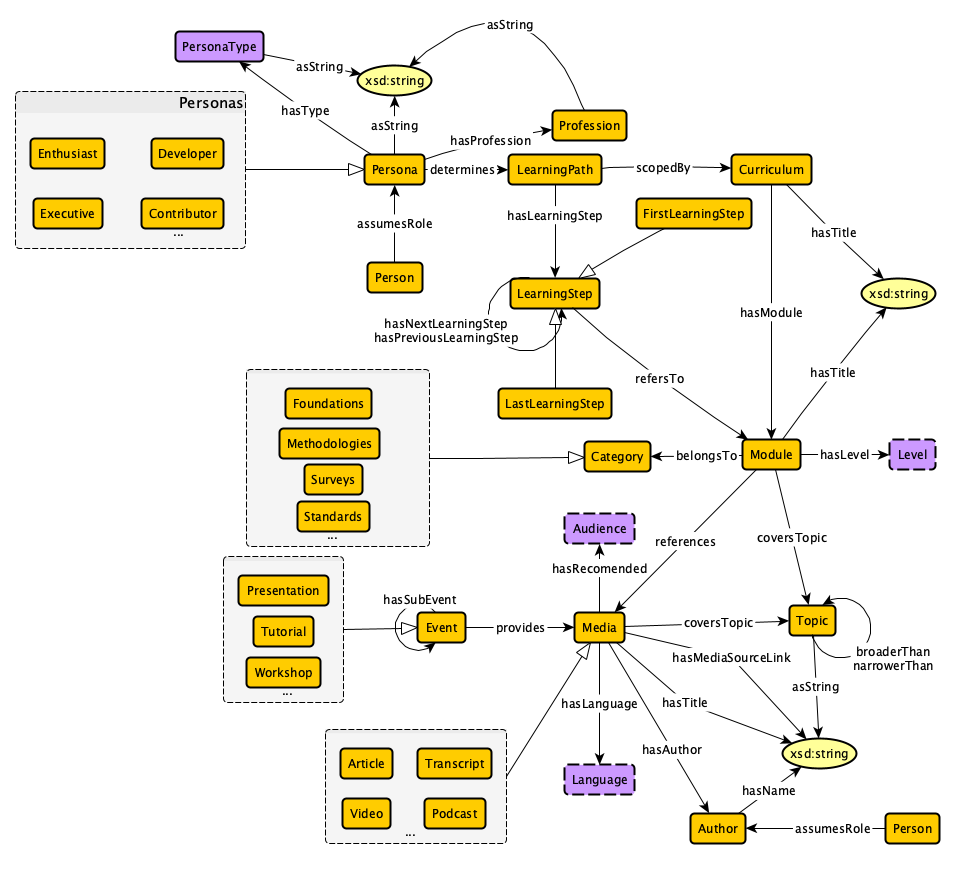}
    \caption{The schema diagram for the overall Curriculum KG Ontology.
    The yellow boxes indicate the classes, the purple ones represent the controlled vocabulary and the yellow ellipses indicate the datatypes tied to those classes. Open arrows indicate SubclassOf relations.}
    \label{fig:schema-diagram}
\end{figure*}

In Figure~\ref{fig:schema-diagram}, we provide a \textbf{schema diagram} of our ontology. A schema diagram is an intuitive, graphical depiction of how concepts are related in the ontology. For each \emph{head\_node-edge-tail\_node} in the diagram, these can be minimally construed as representing axioms of the form $$\textsf{Head} \sqsubseteq \mathord{\geq}0~\textsf{edge.Tail}$$
which we call a structural tautology \cite{owlaxax}. Essentially, it acts as an axiom with no side effects, but helps humans intuit the purpose of a role. The visual syntax is defined as follows. Gold boxes represent classes, purple boxes represent concepts that act as controlled vocabularies\footnote{By this, we mean a class with defined individuals (and no others).}, yellow ellipses represent data types, closed arrows indicate object properties, open arrows indicate subsumption in the direction of the arrow, and grouped boxes are subclasses.

This framework's key concept is the \textsf{LearningPath}, a dynamic series of \textsf{LearningStep}s developed to accommodate different \textsf{Persona}s. These \textsf{Persona}s, so far consisting of developers, enthusiasts, executives, and contributors, direct the learning paths' content and structure. Anyone interested identifies a \textsf{Persona} that fits their role and professional experience, and each \textsf{Persona} shapes the architecture of their respective \textsf{LearningPath}.

The \textsf{FirstLearningStep} is the first step of a learning path, which is composed of a number of steps that are connected to one another. The learning process flows continuously since each step is connected by the \textsf{hasNextLearningStep} and \textsf{hasPreviousLearningStep} properties. These steps are all connected to particular \textsf{Modules} that provide organized educational resources. Instead of existing independently, these modules are arranged within the \textsf{Curriculum}, the top entity representing the comprehensive educational framework that offers a structured educational experience.

Multiple important metadata characteristics define each module. These consist of a title, the level of difficulty, and the \textsf{coversTopic} property, which links the module to specific curriculum subjects.   The topics themselves are organized semantically, allowing for an elaborate topic structure through the use of hierarchical relationships like \textsf{broaderThan} and \textsf{narrowerThan}. For example, the term "Hydrolysis" is recognized as a more narrow problem within the wider field of "Chemistry." This semantic structure enables the efficient navigation of educational materials. 

A \textsf{Module} is a formal unit; it is not identical to the documents, videos, teaching activities, etc. that are used when a student takes the module. These things are to be classified as \textsf{Media}, and a media object can be linked to a \textsf{Module} by the \textsf{references} property.

The \textsf{Event} class is for any educational event that is not directly related to any curriculum, such as an academic conference or workshop. An \textsf{Event} can be a part of a larger event, expressed with the \textsf{hasSubEvent} property. For example, a paper presentation may be part of a conference.

\paragraph{ParticipantRole}
Several aspects of the ontology are dedicated to modeling how \emph{people} interact with the curricula and modules: namely, authorship and as a learner. This involvement -- or participation -- is modeled according to the \textsf{AgentRole} ODP \cite{modl}.\footnote{We note that the AgentRole formulation exists in many forms, but we have sourced in particular these patterns from the MODL pattern library.} In this modeling paradigm, an agent (and usually in our case, that agent is person) is not simply classified as a student, teacher, or so on, but rather, we record that the person \emph{played the role of} a student or teacher at a certain time, or in connection with a certain event. That is, we disconnect what an agent does (or performs) from what that agent is (or who they are). 

Of particular importance for us is the contextualization of that \textsf{ParticipantRole}. Traditionally, the contextualization for a role is the spatiotemporal extent (i.e., an anchoring in space and time). However, for us, we are interested at a more abstract level: ``Who is engaging with specific learning material?'' and ''Who authored a certain piece of material.'' While the date of authorship is important for understanding the usefulness of material (i.e., perhaps in relation to the state of the art), these details are more aptly modeled as part of the media in question.

\begin{figure}[t]
    \centering
    \includegraphics[width=\linewidth]{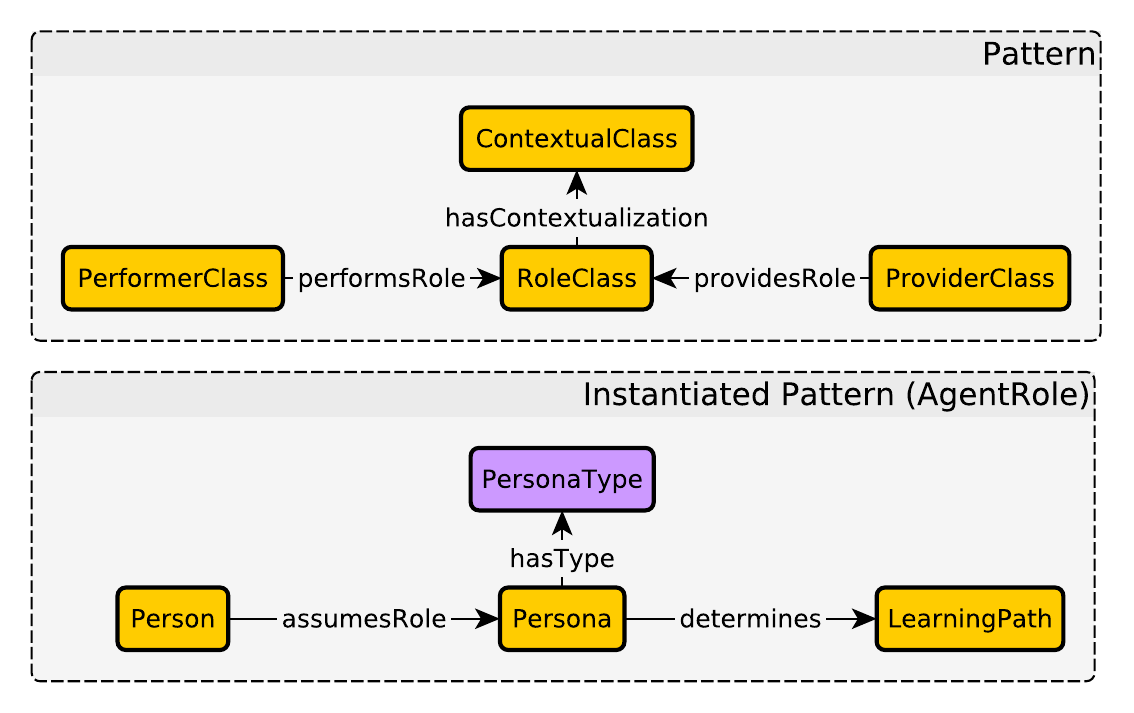}
    \caption{Schematic comparison between the Agent Role ontology design pattern (top) and its instantiated for \textsf{Persona}.}
    \label{fig:agent-role-template}
\end{figure}

Thus, we focus on \textsf{providesParticipantRole}. In our ontology, this is obfuscated however, essentially by focusing on the nature of the contextualization: determination of a suitable learning path and authorship, while structurally this is clear (see Figure~\ref{fig:agent-role-template} We note that an individual of type \textsf{ParticipantRole} is not something generic, like `teacher'. Instead, it represents a specific instance of teaching carried out by one person in one context. For generic roles, we chose to represent them in our ontology as \emph{subclasses}: \textsf{Persona} and \textsf{Author} -- and their subclasses, in turn!

\paragraph{Curriculum}
A \textsf{Curriculum} is a course of study, by which we mean some set of educational material that has been curated for a certain purpose. In higher education, curricula might be described by a syllabus, which includes necessary metadata for engaging with the material: sources (e.g., textbook(s), topics, and a schedule. A prepared course may have learning outcomes and objectives. For now, we largely focus on the topics and the order in which they should be ingested. For the former, these are modeled as \textsf{Module}s and, for the latter, they are modeled as \textsf{LearningPath}s. \textsf{Module}s are organized as bags, which is sufficient for answering questions such as ``What topics does this curriculum cover?'' Axiomatically, we dictate that the \textsf{Curriculum} should be titled, and that (we argue) a curriculum is pointless -- and not much of a curriculum at all -- if there are no modules (and thus topics) that it addresses. While this is a minimal specification, it's important as it acts as a top-level organizational mechanism.

\begin{align}
    \textsf{Curriculum} &\sqsubseteq \exists \textsf{hasTitle.xsd:string} \\
    \textsf{Curriculum} &\sqsubseteq \exists \textsf{hasModule.Module} 
\end{align}

\begin{compactenum}
    \item Every Curriculum has a title represented as a string. 
    \item Every Curriculum has at least one Module.
\end{compactenum}

\paragraph{Person \& Authorship}
A \textsf{Person} can assume different roles, such as an \textsf{Author}, which would be responsible for generating content, or taking on a \textsf{Persona}, which represents an abstracted perspective or set of requirements for an individual. Instead of being only labels, these roles serve to distinguish the kind of contributions or actions that are linked to the person. One may ask, for instance, ``Who developed this module?'' or ``For whom is this series of topics intended?'' Axiomatically, each \textsf{Author} must be a \textsf{Person}, and we consider that each individual's learning path is a reflection of their personal goals.
\begin{align}
    \setcounter{explanations}{\theequation}
    \exists \textsf{assumesAuthorship.Author} &\sqsubseteq \textsf{Person}\\
    \exists \textsf{assumesPersona.Persona}&\sqsubseteq \textsf{Person} \\
    \textsf{Author}\sqsubseteq
    \exists \textsf{hasName.xsd:string}
\end{align}
\begin{compactenum}
    \setcounter{enumi}{\theexplanations}
    \item There exists a Person who assumes the role that is an Author. 
    \item There exists a  Person who assumes a role that is a Persona.
    \item Every author has some name and that is represented as a string.
\end{compactenum}

\paragraph{LearningPath}
A \textsf{LearningPath} is a series of learning steps an individual has to go through in order to follow a certain \textsf{Curriculum} to satisfy a personal educational goal. Just like attending a University class, a certain goal is set for the students that attend, requiring them to go through a set of material in order (path) and therefore accomplish that said goal. It would satisfy the question of ``What Modules do I need to go through to get familiar with Knowledge Graphs in a superficial manner and in what order?'' That path contains certain \textsf{Module}s in sequence, noting that the given \textsf{Persona} will have to complete each one in order to complete their learning objective.

We abbreviate \textsf{hasLearningStep} as \textsf{hLS}.
\begin{align}
    \setcounter{explanations}{\theequation}
    \textsf{LearningPath} &\sqsubseteq \exists \textsf{scopedBy.Curriculum} \\
    \textsf{LearningPath} &\sqsubseteq \exists \textsf{hLS.LearningStep}\\
    \textsf{LearningPath} &\sqsubseteq \exists \textsf{determines}^-\textsf{.Curriculum}
\end{align}
\begin{compactenum} 
    \setcounter{enumi}{\theexplanations}
   \item Every Learning Path is scoped by a Curriculum.
   \item Every Learning Path has at least one Learning Step.
   \item Every Learning Path is determined by at least one Curriculum.
\end{compactenum}

\paragraph{LearningStep}
A \textsf{LearningStep} is an individual step in a \textsf{LearningPath}. This sequence is used to order the \textsf{Module}s in a certain \textsf{Curriculum}, as determined by the relevant \textsf{Persona}. \textsf{LearningStep}s are ordered to provide a mechanism for curation (i.e., to enable an instructor to specify the exact approach), as well as to otherwise encode pre-requisite. In some sense, but which we do not state explicitly, a \textsf{Module} plays the role of a \textsf{LearningStep}, meaning that it is a standalone piece of material, and the context is the order of the \textsf{LearningPath}. The \textsf{LearningPath} is based off the \emph{Semantic Trajectory} \cite{modl} pattern and, as such, includes a \textsf{FirstLearningStep} and \textsf{LastLearningStep}. This concept in particular enables answering simple CQs such as, ``As a developer who wants to work with knowledge graphs in a hands-on way, what module should I go through after 'What is an Ontology?'"

In the axiomatization below, we abbreviate \textsf{hasNextLearningStep} as \textsf{hNLS} (similarly for ``previous''), \textsf{LS} for \textsf{LearningStep} (and prepend F or L for first or last, respectively).
\begin{align}
    \setcounter{explanations}{\theequation}
    \top &\sqsubseteq \mathord{\leq}1~\textsf{hNLS.}\top\\
    \top &\sqsubseteq \mathord{\leq}1~\textsf{hPLS.}\top\\
    \textsf{hPLS.LearningStep} &\sqsubseteq \lnot\textsf{FLS}\\
    \textsf{hNLS.LearningStep} &\sqsubseteq \lnot\textsf{LLS}\\
    \textsf{LearningStep} &\sqsubseteq \mathord{=}1~\textsf{refersTo.Module}
\end{align}
\begin{compactenum}
    \setcounter{enumi}{\theexplanations}
    \item Every learning step has exactly one or no next learning step.
    \item Every learning step has exactly one or no previous learning step.
    \item If a learning step has a previous learning step, then it is not the first learning step in a path.
    \item If a learning step has a next learning step, then it is not the last learning step in a path.
    \item Every learning step refers to exactly one module.
\end{compactenum}

\paragraph{Module}
A \textsf{Module} covers one or more \textsf{Topic}s, has a title to identify it, is assigned a \textsf{Level} to indicate its difficulty, belongs to a specific \textsf{Category} for organization, and may reference relevant \textsf{Media} to support learning. A \textsf{Module} represents an educational material that covers a specific \textsf{Topic}. For example, an article titled ``100+ Data Science Projects in Python for Beginners'' would represent a Module, and the topic of it would be categorized as ``Coding''. It can be categorized into levels, determining a ``difficulty'' measure an audience member can use to satisfy a certain objective while also being prompted to a \textsf{Media} reference such as a book or a video.
\begin{align}
    \setcounter{explanations}{\theequation}
    \textsf{Module} &\sqsubseteq \exists \textsf{coversTopic.Topic} \\
    \textsf{Module} &\sqsubseteq \exists \textsf{hasTitle.xsd:string} \\
    \textsf{Module} &\sqsubseteq \exists \textsf{hasLevel.Level} \\
    \textsf{Module} &\sqsubseteq \exists \textsf{belongsTo.Category} \\
    \textsf{Module} &\sqsubseteq \exists \textsf{references.Media} 
\end{align}
\begin{compactenum}
    \setcounter{enumi}{\theexplanations}
    \item Every Module always covers a Topic. 
    \item Every Module always has a title as a string. 
    \item Every Module always has a level. 
    \item Every Module always belongs to a Category. 
    \item Every Module always references some Media.
\end{compactenum}

\paragraph{Category}
A \textsf{Category} is a subject-matter grouping that organizes relevant modules in a curriculum, helping to meaningfully and conveniently structure the learning content. At least one category must be covered in each module, which gives a foundation for the types of topics covered. We have initially identified four categories; they include \textsf{Foundation}, which provide fundamental ideas and principles; \textsf{Survey}, which covers broad descriptions of a topic or field; \textsf{Methodology}, which describe tools and procedures; and \textsf{Standard}, which offer formal specifications and recommendations. We specifically differentiate from \textsf{Topic} for a more connotative approach. In this case, a \textsf{Category} is a \textsf{Topic} in the broadest sense. Of course, a \textsf{Topic} might cover what a ``standard'' is, but in this case, we are more interested in the fact that the \textsf{Topic} covering, e.g., OWL 2, is about a \textsf{Standard}.
Axiomatically, these are very simple. We do not make any claims to disjointness, and merely require that any particular \textsf{Category} be populated by at least one \textsf{Module}.

\begin{align}
    \setcounter{explanations}{\theequation}
    \textsf{Category} &\sqsubseteq \exists \textsf{hasModule}^-\textsf{.Module}
\end{align}
\begin{compactenum}
    \setcounter{enumi}{\theexplanations}
    \item Every category has some module.
\end{compactenum}

\paragraph{Event}
An \textsf{Event}, within the scope of our ontology, is a structured learning activity that can take several forms, including workshops, tutorials, and presentations, each of which is intended to provide a specific learning modality. \textsf{Sub-events} are related activities that can be included or are part of \textsf{Event}s to allow for a more granular organization of sessions and content. Events can offer access to or otherwise provide \textsf{Media} (e.g, podcasts, articles, videos, and transcripts) that supplement in-person interactions. \textsf{Event}s are frequently tied to the modules they cover. "What type of media is being provided by this Event?" is a question that can be answered using the above framework. From \cite{shimizu2018ontology}, we note that the \textsf{subeventOf} property is effectively a subproperty of \textsf{po-feature}, meaning that we also have meronymous transitivity.
\begin{align}
    \setcounter{explanations}{\theequation}
   \textsf{Event} &\sqsubseteq \mathord{\geq} 0 \ \textsf{hasSubEvent.Event}\\
   \textsf{Event} &\sqsubseteq \exists \textsf{provides.Media}
\end{align}

\begin{compactenum}
    \setcounter{enumi}{\theexplanations}
    \item Every event has zero or more sub-events.
    \item Every event provides some Media.
\end{compactenum}

\paragraph{Persona}
We established a set of \textsf{Personas} to guide progress and evaluate ease of use, covering a range of different kinds of user types, including \textsf{Developers}, \textsf{Instructors}, \textsf{Analysts}, \textsf{Executives}, and 
\textsf{Graduate Students}. By following a customized learning path, each persona allows us to make sure the ontology supports a wide range of technical backgrounds and goals. Respectively, we developed custom \textsf{Learning Paths} for them, including \textsf{Modules} and materials needed for them to go through in order to achieve their separate educational goals. We have specified these strict cardinalities, per below, as each Persona is very granular (i.e., finely and exactly tailored).

\begin{align}
    \setcounter{explanations}{\theequation}
    \textsf{Persona} &\sqsubseteq \mathord{=} 1 \textsf{hasProfession.} \textsf{Profession}\\
    \textsf{Persona} &\sqsubseteq \exists \textsf{hasType.PersonaType}\\
    \textsf{Persona} &\sqsubseteq \mathord{=} 1 \ \textsf{determines.LearningPath}\\
    \textsf{Persona} &\sqsubseteq \exists \textsf{assumesPersona}^-\textsf{.Person}
\end{align}
\begin{compactenum}
    \setcounter{enumi}{\theexplanations}
    \item Every persona has exactly one profession.
    \item Every persona has a type and that type is a persona type.
    \item Every persona determines exactly one learning path.
    \item Every persona is assumed by at least one person.
\end{compactenum}

\paragraph{Topic}
A \textsf{Topic} helps to clarify the main idea covered in a module and assists recognize its purpose. Every topic has a title, and they all act as important content organizers. Because the subjects are arranged in a hierarchy, learners can examine how ideas relate to one another at various levels of complexity. A topic may have both more \textsf{broad} and more \textsf{narrow} related topics. The query, "What are the more advanced topics I should explore after learning about RDF?" is an instance that this model can properly answer.
\begin{align}
    \setcounter{explanations}{\theequation}
     \textsf{Topic} &\sqsubseteq
   \mathord{=} 1\textsf{asString} \textsf{xsd:String}
\end{align}

\begin{compactenum}
    \setcounter{enumi}{\theexplanations}
    \item Every topic is represented by exactly one string value.
\end{compactenum}

\paragraph{Controlled Vocabularies}
The terms \textsf{PersonaType}, \textsf{Level}, \textsf{Audience}, and \textsf{Language} are all controlled vocabularies. This means that the class consists of exactly only the individuals specified. This makes use of the \textbf{Explicit Typing} ODP \cite{modl}, and allows for quick and easy modifications to the ontology without perturbing the overall subsumption hierarchy.

\section{The Curriculum Knowledge Graph}
\label{sec:kg}
In this section, we describe the construction and materialization process of the Curriculum Knowledge Graph (CurrKG) by showcasing how structured curriculum data, which is stored in tables (e.g., CSV files), is turned into a KG using RDF triples. We also present an evaluation using real-world use cases to demonstrate its validity. We utilize competency questions (CQs) to validate the CurrKG and thus demonstrate how the graph effectively represents the required information.

\subsection{Materialization}
\label{ssec:const}
The materialization process consists of translating structured tabular curriculum data into triples conforming to the ontology. While ontology itself provides the definition for conceptual entity models like \textsf{Module}, \textsf{Media}, and \textsf{Persona} (discussed in Section~\ref{sec:description}), the instantiation of these concepts into an RDF graph is achieved using a flexible, code-driven pipeline.

The data used for materialization is drawn from a variety of open and curated sources. It majorly includes The Knowledge Graphs Conference and Community's Open Knowledge Graph Curriculum (Open KGC)\footnote{\url{https://github.com/KGConf/open-kg-curriculum}}, data from National Student Data Corps (NSDC)\footnote{\url{https://nebigdatahub.org/nsdc/}} -- particularly the Ontology Flash Card Series and other consolidated curriculum-related datasets, as well as some internally curated data from our Education Gateway (EduGate) project\footnote{\url{https://edugate.cs.wright.edu/}}. This includes our detailed persona information and learning path structures. All of these datasets collectively are rich and diverse enough to generate instances for key concepts and relationships defined in our ontology.

In abstract, the construction pipeline starts with data ingestion, specifically by inputting a UTF-8 formatted CSV file where each row contains curriculum data entry. These entries are iteratively processed, enabling efficient row-wise iteration and data manipulation. The column headers in the CSV serve as signals for the types of entities and properties to expect in the Python script. We then proceed with the initialization of an RDF graph, using the RDFLib library~\cite{site:rdflib}, with a set of our predefined namespaces for consistent URI and vocabulary, either by loading an existing graph or by creating one from scratch as we proceed from ingestion to triplification. Triplification consists of constructing RDF triples based on the presence and completeness of specific fields in each row of the input data, i.e., whenever a field is missing or empty, the script skips triple creation for that field but continues processing the remainder of the row, preserving semantic structure where available. For example, if both `Module Title` and `Curriculum` are present in a row, all related triples

{\footnotesize
\begin{verbatim}
 (Curriculum_X, hasTitle, Curriculum_Y)
 (Module_Y, hasTitle, Module Title)
 (Curriculum_X, hasModule, Module_Y)
\end{verbatim}
}

are generated. When minting a URI of an entity, we use its identifier in the URI construction by first sanitizing it -- the process of removing all special characters and whitespace replaced by underscores -- and also tagging some available metadata. Finally, we serialize the newly constructed RDF graph into a format such as Turtle (\texttt{.ttl}) \cite{turtle-rec}, which serves as a materialized KG instance of the CurrKG ontology.

Our materialization pipeline and script are adaptable in nature and are not hard-coded to a specific data format or schema. It dynamically responds to the input data structure, handling missing values gracefully and allowing for partial population of entities by instantiating all available individuals of their corresponding ontology classes, even when the data is incomplete. For instance, for classes such as \textsf{Module}, \textsf{Media}, and \textsf{LearningPath}, it creates schema-specific semantic links between individuals, such as \textsf{hasModule} or \textsf{references} and correctly types individuals entities using ontology-defined properties, such as \textsf{hasTitle} or \textsf{asString} as appropriate (see Figure~\ref{fig:schema-diagram} for details). As long as new data follows the expected column headers that correspond to known field mappings, the script can generate all possible RDF triples and materialize a KG instance without modification. This allows CurrKG to be adapted to new datasets, use cases, or domains with minimal engineering effort. The working script and data are available in the CurrKG repository~\cite{repo:currkg}.

\subsection{Evaluation}
\label{ssec:eval}
We evaluate the ontology and KG materialized from it via a diverse set of competency questions (CQs) that cover real-world needs and assess various characteristics of the ontology's quality, including completeness and expressiveness. CQs are natural language queries representing the kinds of questions a KG should be able to answer. They serve as a practical benchmark for assessing the graph or ontology and its alignment with the intended domain~\cite{cqs-keet}.

We then expressed these questions as SPARQL queries~\cite{OntopSWJ}, where we focused on efficiency -- but do not make any guarantees on optimality -- to be executed over the graph using Apache Jena Fuseki Triplestore~\cite{fuseki-docs}. For instance, below are some of the example competency question-query sets used for evaluation; their respective prefixes are listed in Table~\ref{tab:prefix-table}.

\begin{table}[t]
    \small
    \centering
    \begin{tabularx}{\columnwidth}{l|X}
        
        \multicolumn{1}{c}{\textbf{PREFIX}} & \multicolumn{1}{c}{\textbf{URI}}                                     \\ \hline
        edu-r:                                & \textless{}https://edugate.cs.wright.edu/lod/resource/\textgreater{}  \\ 
        edu-ont:                              & \textless{}https://edugate.cs.wright.edu/lod/ontology/\textgreater{}  \\ 
        rdf:                                  & \textless{}http://www.w3.org/1999/02/22-rdf-syntax-ns\#\textgreater{} \\ 
        rdfs:                                 & \textless{}http://www.w3.org/2000/01/rdf-schema\#\textgreater{}       \\ 
        \end{tabularx}
    \caption{Set of Prefixes used in CurrKG}
    \label{tab:prefix-table}
\end{table}

{\footnotesize

CQ1: Which persona is associated with which learning path, and what are the learning steps within that path?
\begin{adjustwidth}{0.25cm}{0.25cm}
\begin{verbatim}
SELECT ?persona ?personaName
?learningPath ?learningStep
?learningStepName ?prevStep
?prevStepName ?nextStep
?nextStepName

WHERE {
 ?persona a edu-ont:Persona ;
  edu-ont:determines 
              ?learningPath ;
  edu-ont:asString
              ?personaName.
 ?learningPath
        edu-ont:hasLearningSteps
              ?learningStep .
 ?learningStep
        edu-ont:asString
              ?learningStepName.
 OPTIONAL {
  ?learningStep
    edu-ont:hasPreviousLearningStep
              ?prevStep .
  ?prevStep
    edu-ont:asString
              ?prevStepName.
  }
  
  OPTIONAL {
   ?learningStep
    edu-ont:hasNextLearningStep
              ?nextStep .
   ?nextStep edu-ont:asString
              ?nextStepName.
   }
   
}
\end{verbatim}
\end{adjustwidth}

This query retrieves all the available \textsf{Persona}, the \textsf{Learning Path} it determines, its corresponding \textsf{Learning Steps}, along with its respective previous and next learning steps if they exist.

CQ2: Which authors have contributed to multiple media resources?

\begin{adjustwidth}{0.25cm}{0.25cm}
\begin{verbatim}
SELECT ?author ?authorName
(COUNT(?media) AS ?count)

WHERE {
 ?media rdf:type edu-ont:Media .
 ?media edu-ont:hasAuthor
              ?author .
 ?author edu-ont:hasName
              ?authorName .
}

GROUP BY ?author ?authorName
HAVING (COUNT(?media) > 1)
ORDER BY DESC(?count)
\end{verbatim}
\end{adjustwidth}
This query retrieves all the available \textsf{Authors} in our KG who have authored more than one \textsf{Media} resource.

CQ3: What topics have the most associated media resources?
\begin{adjustwidth}{0.25cm}{0.25cm}
\begin{verbatim}
SELECT ?topic ?topicName
(COUNT(?media) AS ?mediaCount)

WHERE {
 ?media rdf:type edu-ont:Media .
 ?media edu-ont:coversTopic
              ?topic .
 ?topic edu-ont:asString
              ?topicName .
}

GROUP BY ?topic ?topicName
ORDER BY DESC(?mediaCount)
LIMIT 10
\end{verbatim}
\end{adjustwidth}
This query retrieves the top 10 \textsf{Topics} available in our KG are most frequently covered across all available \textsf{Media}.

CQ4: How many modules belong to each category?
\begin{adjustwidth}{0.25cm}{0.25cm}
\begin{verbatim}
SELECT ?categoryName
(COUNT(?module) AS ?moduleCount)

WHERE {
 ?module a edu-ont:Module ;
    edu-ont:belongsToCategory
            ?category .
 ?category edu-ont:asString
            ?categoryName .
}

GROUP BY ?categoryName
ORDER BY DESC(?moduleCount)
\end{verbatim}
\end{adjustwidth}
This query retrieves the count of all available \textsf{Modules} and the \textsf{Category} they belong to.

CQ5: What are the top 10 most referenced media resources?
\begin{adjustwidth}{0.25cm}{0.25cm}
\begin{verbatim}
SELECT ?media ?mediaTitle
(COUNT(?referencingEntity) AS
                ?referenceCount)
                
WHERE {
 ?referencingEntity
    edu-ont:references ?media .
 ?media edu-ont:hasTitle
              ?mediaTitle.
}

GROUP BY ?media ?mediaTitle
ORDER BY DESC(?referenceCount)
LIMIT 10
\end{verbatim}
\end{adjustwidth}

This query retrieves the top 10 \textsf{Media} that have been referenced the most.
}

Rest of the CQs and their respective SPARQL queries used for evaluation are available in the git repository\footnote{https://github.com/kastle-lab/curriculum-kg}. This approach of executing the SPARQL queries to answer the CQs over materialized CurrKG helped confirm that the ontology sufficiently captured the semantics, hierarchies, and properties of the educational material, and all the entities and relationships were correctly instantiated in materialization \cite{tahsin2024generation,ayadi2024unified}.

We did observe some limitations when the data was incomplete or when key properties  (e.g., \textsf{Event} information) were missing, but the ontology and materialization pipeline's flexibility still allowed for meaningful results to be returned whenever available. This CQ-based evaluation demonstrates and validates the effectiveness of CurrKG's ontology design and its practical implementation in materialization workflow. 

\section{Related Work}
\label{sec:relatedwork}
In this section, important applicable work that offers basic principles of the use of KGs to the organization and customization of educational materials is addressed.   We mention these materials because they support our objective to improve learning through an ontology-based approach. To develop a more adaptable, scalable framework that is customized to each learner's distinct profile, we expand on the approaches used in such studies by examining how they handle entity linking, curriculum representation, and customized learning paths.

\subsection*{A systematic literature review of knowledge graph construction and application in education}
KG construction techniques and their educational applications are carefully reviewed in the systematic review of the literature by Abu-Saliha and Alotaibi \cite{abu2024systematic}. Using a structured evaluation approach, their work creates a collection of knowledge about KG development, integration, and application in educational environments. Through a review of various approaches to entity linking, knowledge representation, and ontology design, their study identifies important frameworks that improve the organization of educational content.   

This review provides helpful details for our ontology-driven KG building, especially when it comes to identifying a structured schema for learning materials and curriculum representation. Their focus on automated knowledge extraction techniques and semantic connections is important to note because it guides our strategy to enhance scalability and flexible learning paths in our KG architecture. In utilizing their research, we improve individualized educational environments by refining our ontology to be more responsive to educational concepts and enable more effective knowledge access.

\subsection*{Towards a Teaching Knowledge Graph for Knowledge Graphs Education}
This work describes a structured framework created especially to enable KG education by integrating key educational components like skills, subjects, courses, instructors, and resources \cite{Ilkou2024TeachingKG}. Their approach is a useful tool for structuring and organizing educational material since it makes use of a higher-order ontology with semantic constraints to guarantee consistency and accessibility. This format makes it possible for learners to access the content in a logical, sequential way, helping to improve understanding and navigation.In contrast to Ilkou's work, which focuses on creating a knowledge graph for learning, our method specifically addresses the difficulties associated with scattered, multimodal instructional resources on many platforms. In order to facilitate smooth progress across educational paths tailored to individual needs, we developed personas and the Curriculum KG ontology, which is linked together and incorporates resources from multiple sources. 

Our work herein was developed in parallel with Ilkou et al. In fact, we were able to cross-pollinate in some cases, leveraging shared experiences. In particular, we build on this approach by modeling individualized learning paths with respect to the personas directly within the KG, capturing unique goals, prior knowledge, and skills to enable truly personalized learning experiences. This enables users with different backgrounds, skill levels, and learning goals to receive tailored educational resources. It improves flexibility and customization, transforming our KG into a dynamic and customized learning assistant in addition to a structured knowledge source.

\subsection*{Knowledge Graph-Based Teacher Support for Learning Material Authoring}
Grévisse et al. \cite{teachersupport-grevisse} describe a system in Knowledge Graph-Based Teacher Support for Learning Material Authoring that uses KGs to assist teachers create educational resources. Their approach, SoLeMiO, uses accessible KGs to create a structured semantic representation while integrating semantic technologies to extract key ideas from lesson plans. In addition to providing semi-automatic classifying to improve reusability across digital learning contexts, this allows the system to identify relevant resources from digital libraries and MOOC platforms. Using current KGs, their strategy mainly aims to enhance educators' content organization and discovery.

To customize educational paths, our work builds on this base but takes a different approach by integrating unique learner personas.  Although Grévisse et al. focus on improving the creation process for educators, we broaden the application of KGs to tailor learning outcomes according to the specific needs of the user.  To ensure that each individual receives material that is in line with their background, objectives, and past knowledge, we model numerous learner profiles within the KG and offer personalized suggestions and flexible learning paths. This distinction makes KGs more dynamic and learner-centric, allowing us to progress beyond content organization topic of personalized learning.

\subsection*{Computer Science Curricula}
Developed by the Accocication for Computing Machinery (ACM), the IEEE Computer Society (IEEE-CS) and the Association for Advancement of Artificial
Intelligence (AAAI), it presents an elaborate model and taxonomy for teaching computer science at the undergraduate level. It organizes the domain into clearly defined Knowledge Areas (KAs), each of which consists of Knowledge Units (KUs) and related Learning Outcomes (LOs) that are arranged according to the levels of intellectual difficulty. Institutions can adapt programs to local needs while keeping them consistent with international standards thanks to this structured approach, which makes it easier to create a modular and competency-based curriculum. By establishing the connections between these taxonomic features and coordinating them with personalities, learning paths, and instructional materials, our Curriculum KG Ontology directly expands upon this framework. We offer an intuitive foundation for dynamically navigating educational content by treating Knowledge Areas and Learning Outcomes as linked graph elements \cite{cs2023}.

\section{Conclusion}
\label{sec:conc}
Ultimately, we have defined an ontology for educational materials and materialized a KG with some available content, using the ontology as a schema.  
By utilizing the concept of \emph{Personas}, we are able to provide customizable learning paths for each member of a specific audience that wishes to get involved with knowledge of this nature. This enables scenarios where individual may learn at their own pace, with respect to their own goals and objectives.

Carefully arranged ontologies like this one allows for further developments and additions with respect to classes, properties, and data. Modularity and reusability are key features that add on to the dynamic nature of the work. Hence, future steps include:
\begin{compactitem}
    \item incorporating the ontology to the features of interactive tools such as the  Interactive Knowledge (InK) Browser\cite{ink-browser}. Serving as an interactive educational platform for the audience members (Personas) to learn according to their personal goals.
    \item Continue to expand the KG with more Personas and Modules that arise throughout our efforts.
    \item Utilizing the organized taxonomies provided in the CS 2023 report as an example for integrating assessment features straight into the Curriculum KG, with special attention to the definition of KAs, KUs, and LOs. Through the compatibility of our ontology with the hierarchical model and complexity-based learning objectives of CS 2023, we hope to facilitate individualized evaluation of learning and automated progress tracking. Workflows for competency-based evaluations will be made possible as a result, assessing comprehension according to semantic approach for ideal results. 
\end{compactitem}

\medskip

\noindent\emph{Acknowledgment.} The author(s) acknowledge support from the National Science Foundation under award \#2333532 ``Proto-OKN Theme 3: An Education Gateway for the Proto-OKN (EduGate). Any opinions, findings, and conclusions or recommendations expressed in this material are those of the author(s) and do not necessarily reflect the views of National Science Foundation or the U.S. Department of Education.


\bibliographystyle{abbrv}
\bibliography{refs}
\end{document}